\documentstyle[12pt,epsf]{article}

\title{
Random-field-driven phase transitions
in the ground state of the
$S=1$ $XXZ$ spin chain
}
\author{Yoshihiro Nishiyama \\
{\it Department of Physics, Faculty of Science,
Okayama University}\\
{\it Okayama 700, Japan}}
\date{(Received \hspace*{50mm})}

\begin{document}
\begin{large}

\maketitle



\section*{Abstract}
Ground-state of the $S=1$ $XXZ$ spin chain
under the influence of the random magnetic field
is studied by means of the exact-diagonalization method.
The $S=1/2$ counterpart has been investigated extensively so far.
The easy-plane area, including the Haldane and the $XY$ phases,
is considered.
The area suffers significantly from the magnitude of the constituent spin.
Destruction of the Haldane state is observed at a critical
strength of the random field, which is comparable to the 
magnitude of the Haldane gap.
This transition is characterized by the disappearance of the
string order.
The $XY$ region continues until at a critical randomness,
at which a transition of the KT-universality class occurs.
These features are contrasted with those
of the $S=1/2$ counterpart.





\section{Introduction}
\label{section1}

In {\it quantum} statistical mechanics,
randomnesses still remain non-trivial,
even though they do not introduce
any competing interactions.
The essence of the integer quantum hole effect,
for example, is attributed to
the randomness-driven localization-delocalization transition
under the presence of high magnetic field.
Although this description is simply of the one-body picture,
it is very hard to deal with this problem.
The effect of randomness for the quantum {\it many-body} system
would be more complicated, and
has not been investigated very well.
In order to consider the effect of randomness for such the systems
as the Mott insulator and
the superconductor, for example,
it is significant to take into account of the many-body interactions
\cite{Fisher89}.

In one dimension, some analytic treatments are available:
The random $S=1/2$ $XXZ$ model was investigated with the 
bosonization technique \cite{Dorty92} and the real-space decimation method
\cite{Fisher94}.
The real-space decimation method yields comprehensive 
understanding of the random transverse-field Ising chain as well
\cite{Fisher92,Fisher95}.
The analytical predictions of these theories are confirmed numerically
\cite{Nagaosa87,Haas93,Runge94,Young96}.
In the course of the above studies,
one could resort to very nice and special mathematical 
characteristics.
That is, the $S=1/2$ chain is described in
the language of the bosonization formalism, while
the transverse-field Ising chain
is related to
the free-fermion model with the Jordan-Wigner transformation.
Because both the resultant models are understood very well,
the effect of the randomness could be considered to 
a certain extent.

Recently, more general systems, which are possibly missing
such the nice characteristics as above,
have been attracting considerable attention
\cite{Boechat96,Hyman96,Yang96,Hyman97,Monthus97,Hida97,%
Nishiyama97,Todo97}.
It is suggestive that
the real-space decimation method fails for the cases other than $S=1/2$
\cite{Fisher94,Hyman97}.
This is somehow consistent with the Haldane conjecture 
\cite{Haldane83}, which states
that the quantum magnet is affected by the magnitude of the spin 
quantitatively.
According to his conjecture, the ground state of the $S=1$ chain
is massive.
Hence, in this case, the technique 
\cite{Dorty92,Giamarchi87,Giamarchi88} based on the
bosonization formalism fails
as well.
Few studies have been reported for the $S=1$ magnet so far
\cite{Boechat96,Yang96,Hyman97,Monthus97,Nishiyama97}.
These studies are concerned in the bond-random
$S=1$ Heisenberg magnet.

Here, we study the $S=1$ $XXZ$ model with the random magnetic field,
\begin{equation}
{\cal H}=\sum_{i=1}^L 
\left(
S^x_i S^x_{i+1} + S^y_i S^y_{i+1} 
+ \lambda S^z_i S^z_{i+1}
+ H_i S^z_i
\right).
\label{Hamiltonian}
\end{equation}
The operators $\{ {\bf S}_i \}$ denote the $S=1$ spin operators
acting on the site $i$.
The random field $\{ H_i \}$ distributes uniformly
over the range $[ -\sqrt{3} \Delta , \sqrt{3} \Delta ]$.
Therefore the mean deviation is given by 
$\sqrt{ [ H_i^2 ]_{\rm av}}=\Delta$.

Under the absence of the random field $\Delta=0$,
the ground state phase diagram of the system 
(\ref{Hamiltonian}) is known \cite{Schulz86,Sakai90}:
(It is noteworthy that
the phase diagram is different qualitatively from that
of the $S=1/2$ chain, which is reviewed in the next section.)
In the region $\lambda<-1$, the ground state is ferromagnetic.
In the region $-1<\lambda<0$, the ground state is 
in the $XY$ phase.
Therefore, in this area,
the ground-state magnetic correlation decays obeying the power law,
\begin{equation}
\langle S^x_i S^x_j \rangle
\sim
1/|i-j|^{\eta_{S=1}}.
\label{eta_S1}
\end{equation}
The exponent $\eta_{S=1}$ is speculated to 
\cite{Alcaraz92} be described by the formula,
\begin{equation}
\eta_{S=1}=\frac{\pi-\cos^{-1}\lambda}{2\pi}.
\end{equation}
In the region $0<\lambda<1.2\cdots$, the Haldane phase extends.
In the Haldane phase, The ground state is massive,
and the magnetic correlation decays exponentially,
\begin{equation}
\langle S^x_i S^x_j \rangle \sim
{\rm e}^{-|i-j|/\xi}.
\label{exponential}
\end{equation}
At the isotropic point $\lambda=1$, the Haldane gap $\Delta E_{\rm triplet}$
and
the correlation length $\xi$ are
estimated numerically \cite{Golinelli94};
$\Delta E_{\rm triplet}=0.41049(2)$ and $\xi=6.2$, respectively.
The antiferromagnetic phase appears in the remaining region
$1.2\cdots<\lambda$.
Among the above four phases,
the antiferromagnetic and the ferromagnetic phases
are of rather classical nature.
Hence, in this paper, we devote ourselves to the easy-plane region,
namely, the Haldane and the $XY$ phases.
Both phases are 
realized precisely due to 
the quantum effect, and the influence of the randomness is 
unclear.

The present paper is organized as follows.
First, we review the results obtained for the $S=1/2$ model
in the next section.
In Section \ref{section3}, our numerical results 
for the $S=1$ chain (\ref{Hamiltonian}) are presented.
First, we focus on the transition from the Haldane phase to
the localization phase at the isotropic point $\lambda=1$.
For the first time, we observed the onset of the localization 
transition from the Haldane phase at $\Delta_{\rm c} = 0.49\pm0.15$,
with the critical exponent $\nu=3.4\pm2.2$.
The transition is characterized as the disappearance of the string correlation.
In the remaining part, we concentrate on the $XY$ region $-1<\lambda<0$.
We find that the $XY$ phase persists considerably
against the randomness, and the randomness-riven phase transition is of
the KT universality class.
In the last section, we summarize these results, and
make a comparison between ours and those of $S=1/2$.

\section{Review ---
the ground-state phase diagram and the criticality
of the $S=1/2$ $XXZ$ chain with the random magnetic field}
\label{section2}

Here, we summarize the ground-state property of the
the Hamiltonian (\ref{Hamiltonian})
in the case of $S=1/2$.
First, we explain
the phase-diagram without randomness
\cite{Yang66a,Yang66b}:
the $XY$ phase spans the whole easy-plane region $-1 < \lambda < 1$.
In this $XY$ phase, the critical exponent $\eta_{S=1/2}$
is evaluated exactly 
\cite{Baxter72} as the function of the anisotropy $\lambda$,
\begin{equation}
\eta_{S=1/2}=\frac{\pi-\cos^{-1}\lambda}{\pi}.
\label{eta_S12}
\end{equation}
Beside this $XY$ phase, the ferromagnetic phase appears in
$\lambda<-1$, while the antiferromagnetic phase extends in
$1<\lambda$.

The ground-state
phase diagram in the presence of the random field 
is depicted in 
Fig. \ref{phase_diagram_S12}.
The diagram is determined with various means,
the bosonization technique \cite{Dorty92},
real-space decimation method \cite{Fisher94} and the
exact-diagonalization method \cite{Runge94}.

If a ground state is critical and the critical exponent
$\eta$ is known, Harris's criterion 
\cite{Harris74} tells whether randomness
is a relevant perturbation or not.
The Harris criterion in the present case \cite{Dorty92}
states that the random-field perturbation is relevant
(irrelevant), if the exponent is $\eta>1/3$ ($\eta<1/3$);
and so, $\eta_{\rm c}=1/3$.
Note that this criterion together with the formula (\ref{eta_S12})
yields the conclusion consistent to the phase diagram shown in Fig. 
\ref{phase_diagram_S12};
the $XY$ phase is stable in the anisotropy range
$-1<\lambda<-0.5$, whereas it is unstable in the other range $-0.5<\lambda<1$.

The criterion, however, is concerned only with
the vicinity of the line $\Delta=0$.
Hence, we must resort to numerical simulations in order to
investigate the region of the finite randomness $\Delta>0$.
The phase diagram of the full range of the randomness
was explored by means of the numerical simulation by Runge and Zimanyi
\cite{Runge94}.
In our case $S=1$, there also appears the Haldane phase, which is 
driven out of a certain criticality.
Therefore, no rigorous criterion such as Harris's is
available even in the vicinity of the pure line.
These are the reasons why we managed the numerical simulation.

Lastly, we mention physical interpretation of the phase diagram.
The $S=1/2$ magnet is transformed with the Jordan-Wigner transformation
to the spinless-fermion system,
\begin{equation}
{\cal H}_{\rm sf} = \sum_i
\left\{
0.5( c^\dagger_i c_{i+1}+{\rm h.c.} )
+\lambda(c^\dagger_i c_i -1/2)(c^\dagger_{i+1}c_{i+1} -1/2)
+0.5H_ic^\dagger_i c_i
\right\}.
\end{equation}
In consequence, we see that the parameter $\lambda$ controls
the strength of the repulsion among the particles,
and the randomness $\Delta$ enters as the potential randomness.
In general, the particles become stable against the randomness
as the cohesive force is strengthened;
the tendency is quite consistent with the phase
diagram in Fig. 
\ref{phase_diagram_S12}.

\section{Numerical results}
\label{section3}

In this section, we investigate the ground state of the
model
(\ref{Hamiltonian})
by means of the exact-diagonalization method.
The results for 
the Haldane and the $XY$ phases are
presented separately
in respective subsections.
In the subsections \ref{section3_1} and \ref{section3_2},
each data is averaged over $128$ and $160$ samples, respectively.
The diagonalization was performed
within the sub-space of the quantum number
$\sum_i^L S^z_i=0$.
This sub-space is the most relevant to the ground-state:
In fact, at $\Delta=0$, the ground state belongs to this
subspace.
This would not be violated essentially even in the presence of finite
randomnesses $\Delta>0$ because of the symmetry of the Hamiltonian
(\ref{Hamiltonian}) under the transformation $z \leftrightarrow -z$.

\subsection{Random-field driven phase transition from
the Haldane phase}
\label{section3_1}

As is explained in Section \ref{section1},
in the region $0<\lambda<1.2\cdots$ ($\Delta=0$), the Haldane phase 
is realized.
At the isotropic point $\lambda=1$,
the magnitude of the Haldane gap becomes maximal.
Here, we concentrate on this point $\lambda=1$.

Some might wonder that the Haldane phase and the localization
(random-field) phase are equivalent, because 
they both exhibit short-range magnetic correlations.
Here, however, we show for the first time
that these phases are distinguishable, in other words,
a critical point separates these phases.

In order to discriminate the phases,
we utilized
the string
correlation \cite{den89,Tasaki91},
\begin{equation}
{\cal O}^x_{\rm string}(j-i)=
\left[
\left\langle 
S^x_i {\rm e}^{{\rm i}\pi\sum_{k=i}^{j-1} S^x_k} S^x_j 
\right\rangle
\right]_{\rm av}.
\label{string}
\end{equation}
Here, the bracket $\langle \ \rangle$ denotes the ground-state
expectation value, and
the symbol $[\ ]_{\rm av}$ stands for the random average.
The string order develops 
\cite{Girvin89} in the ground state of the
Haldane phase, although the usual magnetic correlation is short-ranged;
see eq. (\ref{exponential}).

Because of the presence of the random field parallel to the
$z$ axis, the $z$-component correlation is suffered from 
the random-field disturbance directly.
Hence, we have evaluated the $x$-component correlation,
as is defined in eq. (\ref{string}).
(In the numerical simulation, such a off-diagonal operator is
more difficult to treat than the diagonal one.
This difficulty limited the maximal system size and the
number of samples available.)

In Fig. \ref{u_1}, we plotted the Binder parameter \cite{Binder81}
associated
with the string correlation (\ref{string}),
\begin{equation}
U_{\rm string} = 1 - \frac{ \langle M_{\rm string}^4 \rangle}
                          { 3 \langle M_{\rm string}^2 \rangle^2 },
\label{Binder}
\end{equation}
where the string-order magnetization is given by,
\begin{equation}
M_{\rm string}=\sum_{i=1}^L S^x_i {\rm e}^{{\rm i}\pi\sum_{k=1}^{i-1}S^x_k}.
\end{equation}
As the system size is increased,
the Binder parameter approaches to $2/3$, if the corresponding
order, namely, the string order, develops.
On the other hand, it vanishes in the disorder region,
as the system size is enlarged.
It is system-size invariant at the critical point.
Actually, we see in Fig. \ref{u_1},
that in the region $\Delta<0.5$ the Binder parameter grows
through enlarging the system size,
while it is suppressed in the opposite area $0.5<\Delta$.
In consequence, 
we see that the Haldane phase persists against the randomness
up to $\Delta_{\rm c}\sim0.5$, whereas it is disturbed by the random field
in the region $\Delta>0.5$.

In order to investigate the criticality more precisely,
we employed the finite-size-scaling technique.
The finite-size-scaling theory tells that a
dimensionless quantity such as the Binder parameter 
should be described by the formula,
\begin{equation}
U(L)=\tilde{U}(L/\xi).
\label{Binder_scaling}
\end{equation}
The correlation length diverges in the
vicinity of the critical point $\Delta_{\rm c}$
in the form,
\begin{equation}
\xi=|\Delta -\Delta_{\rm c}|^{-\nu},
\label{power_law}
\end{equation}
where $\nu$ is the correlation-length critical exponent.
As a consequence,
the $\left[(\Delta-\Delta_{\rm c})L^{1/\nu}\right]$-$U$
plots should form an universal
curve irrespective to the system size $L$.
In other words, 
the scaling parameters, such as $\Delta_{\rm c}$ and
$\nu$, are chosen so that the plots, the so-called scaling plots,
align all along a curve.
The degree of the alignment is measured in the way explained in 
Appendix \ref{appendix}.

The scaling plot is presented 
in Fig. \ref{pw_scaling_1}.
The plot determined the critical randomness $\Delta_{\rm c}=0.49\pm0.15$ and
the critical exponent $\nu=3.4\pm2.2$.
We find that the magnitude of the critical point 
is comparable with the magnetic excitation gap 
$\Delta E_{\rm triplet}=0.41049(2)$ \cite{Golinelli94}
at the condition
$\Delta=0$ and $\lambda=1$.

\subsection{Random-field-driven transition from the $XY$ phase}
\label{section3_2}

Here we investigate the $XY$ phase $-1 < \lambda < 0$.
The $XY$ phase is characterized by the presence of
the spin stiffness \cite{Fisher73},
\begin{equation}
\rho_{\rm s}=L
\left[
\left\langle
 \frac{\partial^2 E_0(\theta)}{\partial \theta^2}
\right\rangle
\right]_{\rm av}.
\label{stiffness}
\end{equation}
The symbol $E_0(\theta)$ denotes the ground state energy, and
the angle $\theta$ is associated with the gauge twist of the boundary
condition;
$S^+_L S^-_1 \to {\rm e}^{{\rm i}\theta} S^+_L S^-_1$.
In other words,
the spin stiffness measures the long-range coherence of the gauge symmetry,
and thus it is related to the superfluid density in the context of 
the electron system.
The spin stiffness should vanish in the localization 
(random-field) phase, because the global coherence is
destructed there.

In Fig. \ref{sf_mt5},
we plotted the spin stiffness against the randomness.
The stiffness is suppressed, as the randomness is strengthened.
In the weak-random region $\Delta<1$,
the stiffness hardly changes with respect
to the system size.
Hence, we observe in this area, the stiffness remains finite
in the thermodynamic limit;
that is, in this region, the $XY$ phase is realized.
In the area $\Delta>1$,
however, the stiffness vanishes rapidly through
enlarging the system size.
Each spin is enforced to point towards the random-field direction;
the localization phase is realized there.
To summarize, we see that the $XY$ phase persists up to the critical 
randomness $\Delta_{\rm c}\sim1$,
and beyond the threshold the ground state becomes
localized.

We explore the criticality with the finite-size-scaling method.
Because the stiffness is system-size invariant at the critical point,
the stiffness is described by the formula,
\begin{equation}
\rho_{\rm s}=\tilde{\rho_{\rm s}}(L/\xi).
\label{sf_scaling}
\end{equation}
(Note that it is identical with the form for the Binder 
parameter (\ref{Binder_scaling}).)
The correlation length $\xi$ would obey 
the KT type form,
\begin{equation}
\xi\sim{\rm e}^{A/\sqrt{\Delta-\Delta_{\rm c}}},
\label{xi_KT}
\end{equation}
because a finite critical region extends as is shown above.

The scaling data for $\lambda=-0.5$ is presented in Fig. 
\ref{kt_scaling_mt5}.
This plot yields the estimates $\Delta_{\rm c}=0.85\pm0.30$
and $A\sim0.65$.
(For the details of the analysis, refer to Appendix \ref{appendix}.)

In order to confirm the above result,
we show another analysis for determining
the transition point.
As is reviewed in Section \ref{section2},
the Harris criterion insists that
the correlation decays with the exponent $\eta_{\rm c}=1/3$
at the localization-delocalization transition point.
In order to adopt this criterion,
we plotted the log-log plot of the square of the
normalized staggered magnetization,
\begin{equation}
M_{\rm stg}^2/L^2=1/L^2
\left(
\sum_i (-1)^i S^x_i
\right)^2\sim1/L^\eta,
\label{staggered}
\end{equation}
for various randomnesses in Fig. \ref{corr_mt5}.
We observe that the above criterion yields the estimate $\Delta_{\rm c}\sim 1$.
This conclusion is consistent with the former estimate
$\Delta_{\rm c}=0.85\pm0.30$.
The agreement would be satisfactory.

Secondly, we present the results at the condition $\lambda=0$;
see Figs. \ref{sf_0}-\ref{corr_0}.
The results resemble those of $\lambda=-0.5$:
The scaling analysis for the spin stiffness yields the
critical point $\Delta{\rm c}=0.87\pm0.30$ (see Fig. \ref{kt_scaling_0}),
whereas the Harris criterion suggests that $\Delta{\rm c}\sim1$
(see Fig. \ref{corr_0}).
Again, both estimates are in good agreement.

Lastly, we show the results for $\lambda=-1$;
see Figs. \ref{sf_m1}-\ref{corr_m1}.
(We found that in the range $\Delta>0.6$ of $L=14$,
diagonalization convergence is hardly achieved for
some random samples.
Hence, the corresponding data are missing in these figures.)
We observe in Fig. \ref{sf_m1} that solely at $\Delta=0$,
$XY$ state is realized, and for the region other than that,
the localization phase is realized.
In fact, we find that the data is scaled well under the assumption
of the formula
(\ref{power_law}); see Fig. \ref{pw_scaling_m1}.
The analysis gives the criticality $\Delta_{\rm c}=-0.13\pm0.10$ and
$\nu=0.70\pm0.10$.
The negative value of the randomness is apparently non-sense.
Considering a certain statistical error,
we suspect that the transition point is located at
$\Delta\sim0$.
The Harris criterion, see Fig. \ref{corr_m1},
confirms this conclusion:
In the figure, it is shown clearly that the correlation is suppressed
owing to the randomness significantly.
That is, all the data are curved convexly.
This fact suggests that the correlation decays exponentially
even for the infinitesimal randomness.

\section{Summary and discussions}
\label{section4}

We have investigated the $S=1$ $XXZ$ chain with the random magnetic field
(\ref{Hamiltonian}) by means of the exact-diagonalization method.
To summarize our numerical results presented in Section \ref{section3},
we depicted a drawing of the ground-state phase diagram 
as in Fig. \ref{phase_diagram_S1}.
This diagram should be contrasted to that of the $S=1/2$ 
model shown in Fig. \ref{phase_diagram_S12}.

First, we discuss the Haldane phase ($\lambda=1$).
Although both the Haldane and the random-field phase
are magnetically disordered, these phases are separated
distinctively by the phase boundary.
The Haldane phase is characterized by the presence of the string order
(\ref{string}),
whereas the localization phase is missing the order.
We found that 
the phase boundary is located
at the random-field strength $\Delta_{\rm c}=0.49\pm0.15$.
It is noteworthy that this critical value is comparable 
with the magnitude of the Haldane gap, $\Delta E_{\rm triplet}=0.41049(2)$
\cite{Golinelli94}.
Moreover, 
we obtained 
the critical exponent
$\nu=3.4\pm2.2$ with the finite-size-scaling analysis.
At present, little is reported on the ground-state criticality
driven by {\it randomness} on related systems:
In the case of $S=1/2$ at the isotropic point $\lambda=1$,
the localization-delocalization transition occurs at $\Delta_{\rm c}=0$
with the exponent $\nu=1$ \cite{Fisher94,Runge94}.
Our universality class would not be identical to this,
and possibly unique.

Lastly, we mention the $XY$ phase ($-1<\lambda<0$).
The phase boundary is determined both by the scaling plot
of the spin stiffness and the Harris criterion on the correlation-%
decay rate.
Both estimates are in good agreement.
We observed that the whole region $-1<\lambda<0$ is stable against the
randomness, and the
phase boundary extends up to the random strength $\Delta\sim0.9$.
Note that in the case of $S=1/2$, the $XY$ phase is
stable only in the range $-1<\lambda<-0.5$, and furthermore
the maximal critical randomness is $\sim 0.12$ at most.
These maximal critical randomnesses
are far apart.
Even if we normalize the energy unit in terms of the magnitude of the 
respective spins,
the discrepancy between these estimates
is not resolved completely.
We conjecture that through enlarging the magnitude of the constituent spin,
classical nature emerges very rapidly.
That is to say, for a large-spin chain,
the ground-state magnetism is very stable
in respect to the randomness.
This consequence is quite reasonable, because
classical magnetic order survives even in the presence of
the random field at the ground state.

We have not dealt with the region a bit exceeding the point $\lambda=0$.
At this point ($\Delta=0$), the KT transition takes place;
The energy gap starts to open {\it very slowly}.
Hence, severe correction 
\cite{Malvezzi95} to the finite-size scaling emerges.
This difficulty is remained to be solved for the future.

\section*{Acknowledgement}
Our computer programs
are partly based on the subroutine package ``TITPACK Ver. 2''
coded by Professor H. Nishimori.
Numerical simulations were performed on the parallel
supercomputers FUJITSU VPP700/56
of the computer center, Kyushu university.

\appendix

\section{Details of the present scaling analyses}
\label{appendix}

We explain the details of our finite-size-scaling analyses,
which we managed in Section \ref{section3}
in order to estimate the transition point $\Delta_{\rm c}$,
the exponent $\nu$ in eq. (\ref{power_law}), and
the coefficient $A$ in eq. (\ref{xi_KT}).
We adjusted these scaling parameters
so that the scaled data, such as those shown in
Figs. \ref{pw_scaling_1} and \ref{kt_scaling_mt5},
form a curve irrespective of the system sizes.
In order to see quantitatively to what extent these data align,
we employ the ``local linearity function'' $S$
defined by Kawashima and Ito \cite{Kawashima93}:
Suppose
a set of the data points $\{ (x_i,y_i) \}$ with the errorbar
$\{ d_i(=\delta y_i) \}$, which
we number so that $x_i<x_{i+1}$ may
hold for $i=1,2,\cdots,n-1$.
For this data set, the local-linearity function is defined as
\begin{equation}
S=\sum_{i=2}^{n-1} 
w(x_i,y_i,d_i|x_{i-1},y_{i-1},d_{i-1},x_{i+1},y_{i+1},d_{i+1}).
\label{linear}
\end{equation}
The quantity $w(x_j,y_j,d_j|x_i,y_i,d_i,x_k,y_k,d_k)$
is given by
\begin{equation}
w=
\left(
\frac
{y_j-\bar{y}}
{\Delta}
\right)^2 \ ,
\end{equation}
where
\begin{equation}
\bar{y}=\frac{(x_k-x_j)y_i-(x_i-x_j)y_k}
{x_k-x_i}
\end{equation}
and
\begin{equation}
\Delta^2=d^2_j+
\left(
\frac{x_k-x_j}{x_k-x_i} d_i
\right)^2
+
\left(
\frac{x_i-x_j}{x_k-x_i}d_k
\right)^2.
\end{equation}
In other words, the numerator $y_j-\bar{y}$ 
denotes the deviation of the point 
$(x_j,y_j)$
from the line passing
two points $(x_i,y_i)$ and $(x_k,y_k)$, and
the denominator $\Delta$ stands for the statistical error
of $(y_i-\bar{y})$.
And so, $w=((y_i-\bar{y})/\Delta)^2$ shows a degree to what extent
these three points align.
The advantage in the above analysis is as follows:
In the conventional least square fitting, we need to assume
some particular fitting function. The assumption
which function we use causes systematic error.
Note that in the present analysis, we do not have to assume
any fitting functions.

The scaling parameters are determined so as to minimize the
local-linearity function.
The error margin of the scaling parameter is hard to estimate.
It is concerned with both the statistical error and the correction to the
finite size scaling.
The former error can be estimated through considering the 
statistical error of the function $S$.
This function has a relative error of the order $1/\sqrt{n-2}$.
(The local-linearity function is of the order $(n-2)$,
and the error is given by $\sqrt{n-2}$.)
The correction to the finite-size scaling,
on the contrary, is quite difficult to determine.
In the present study, in particular, the KT transition
occurs. 
In that case, in general, the correction to the finite-size scaling
emerges severly.
Here, we consider only the statistical error in order
to estimate the error margins of the scaling parameters.
As the number of the data points $n$ is increased,
the statistical error of $S$ would be reduced.
The corrections to the finite-size scaling
might increase instead.
In the present analyses, we used twenty data in the vicinity of the
transition point $\Delta_{\rm c}$.
An example of the plot $S$ is shown in Fig. \ref{linearity}.
We observe the minimum at $\nu\approx 1.5$.
Taking into account of the statistical error,
we estimated the critical exponent as $\nu=3.4\pm2.2$.



\begin{figure}[htbp]
\begin{center}\leavevmode
\epsfxsize=17cm
\epsfbox{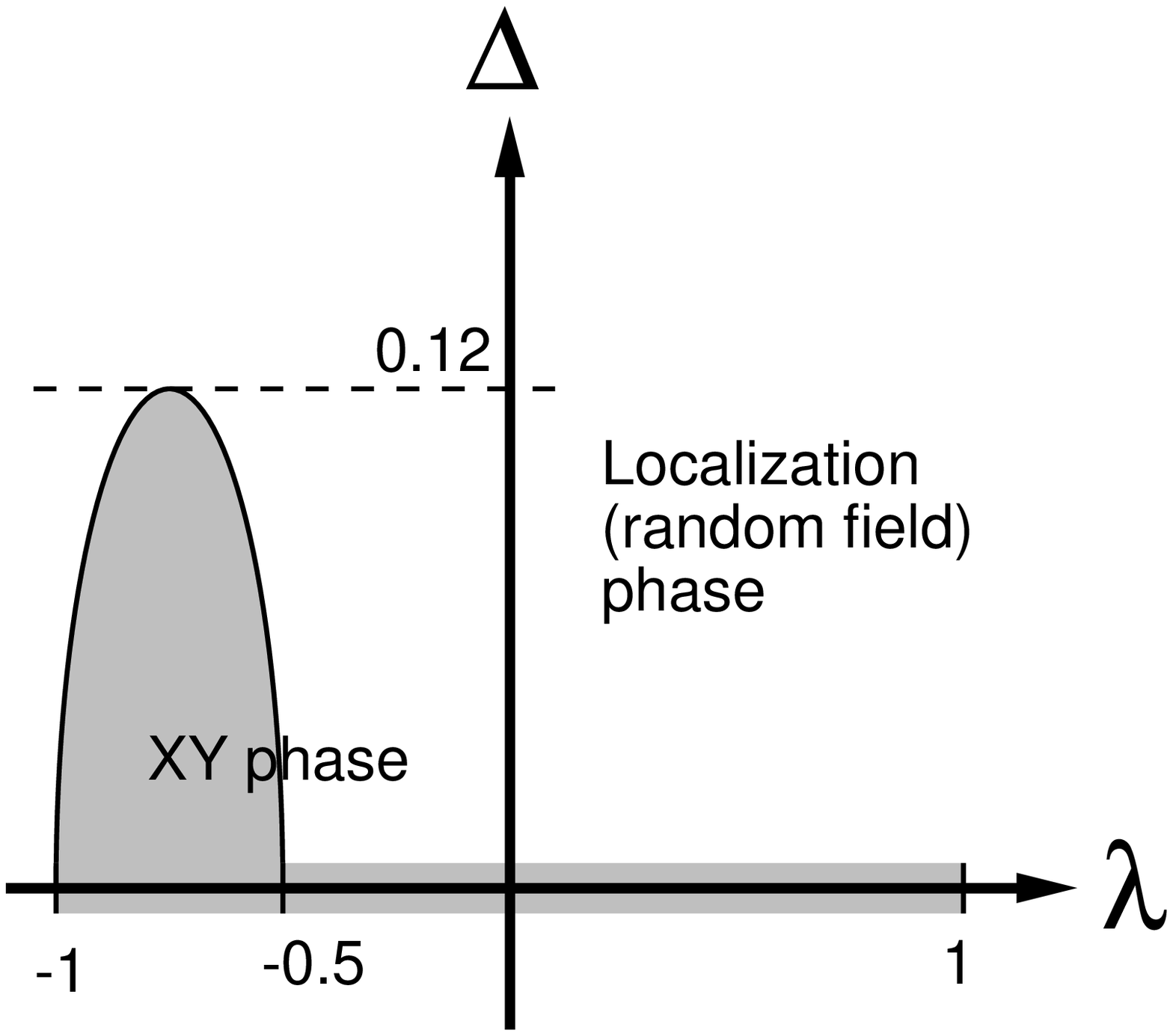}
\end{center}
\caption{
The ground-state phase diagram of the $S=1/2$ $XXZ$ model 
({\protect \ref{Hamiltonian}})
under the presence of the random magnetic field $\Delta$.
Here, we have
referred to the figure appearing in the article {\protect \cite{Runge94}}.}
\label{phase_diagram_S12}
\end{figure}

\begin{figure}[htbp]
\begin{center}\leavevmode
\epsfxsize=17cm
\epsfbox{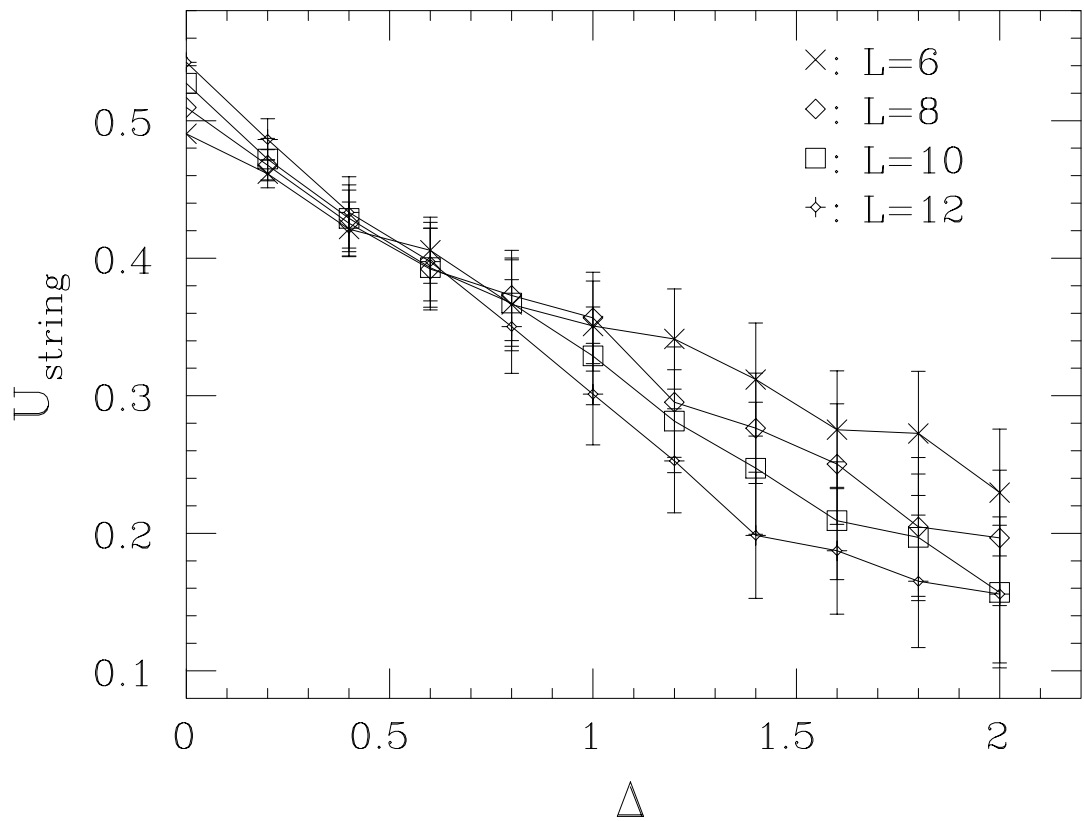}
\end{center}
\caption{
Plot of the Binder parameter ({\protect \ref{Binder}}) associated with
the string order ({\protect \ref{string}}) with the randomness $\Delta$ varied
at the isotropic point $\lambda=1$.
The system-size-invariant point
indicates the location of the critical point.}
\label{u_1}
\end{figure}

\begin{figure}[htbp]
\begin{center}\leavevmode
\epsfxsize=17cm
\epsfbox{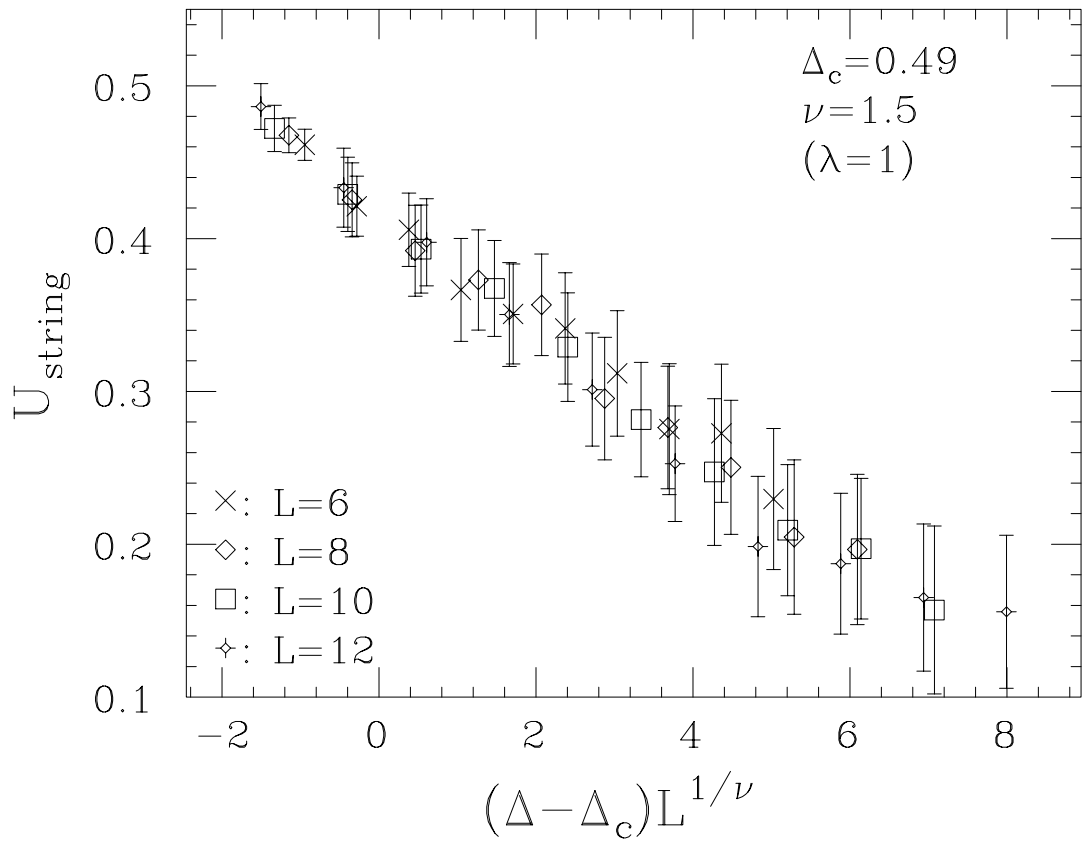}
\end{center}
\caption{
The scaling plot of the data shown in Fig. {\protect \ref{u_1}}.
This plot yields the estimates $\Delta_{\rm c}=0.49\pm0.15$ and 
$\nu=3.4\pm2.2$.}
\label{pw_scaling_1}
\end{figure}

\begin{figure}[htbp]
\begin{center}\leavevmode
\epsfxsize=17cm
\epsfbox{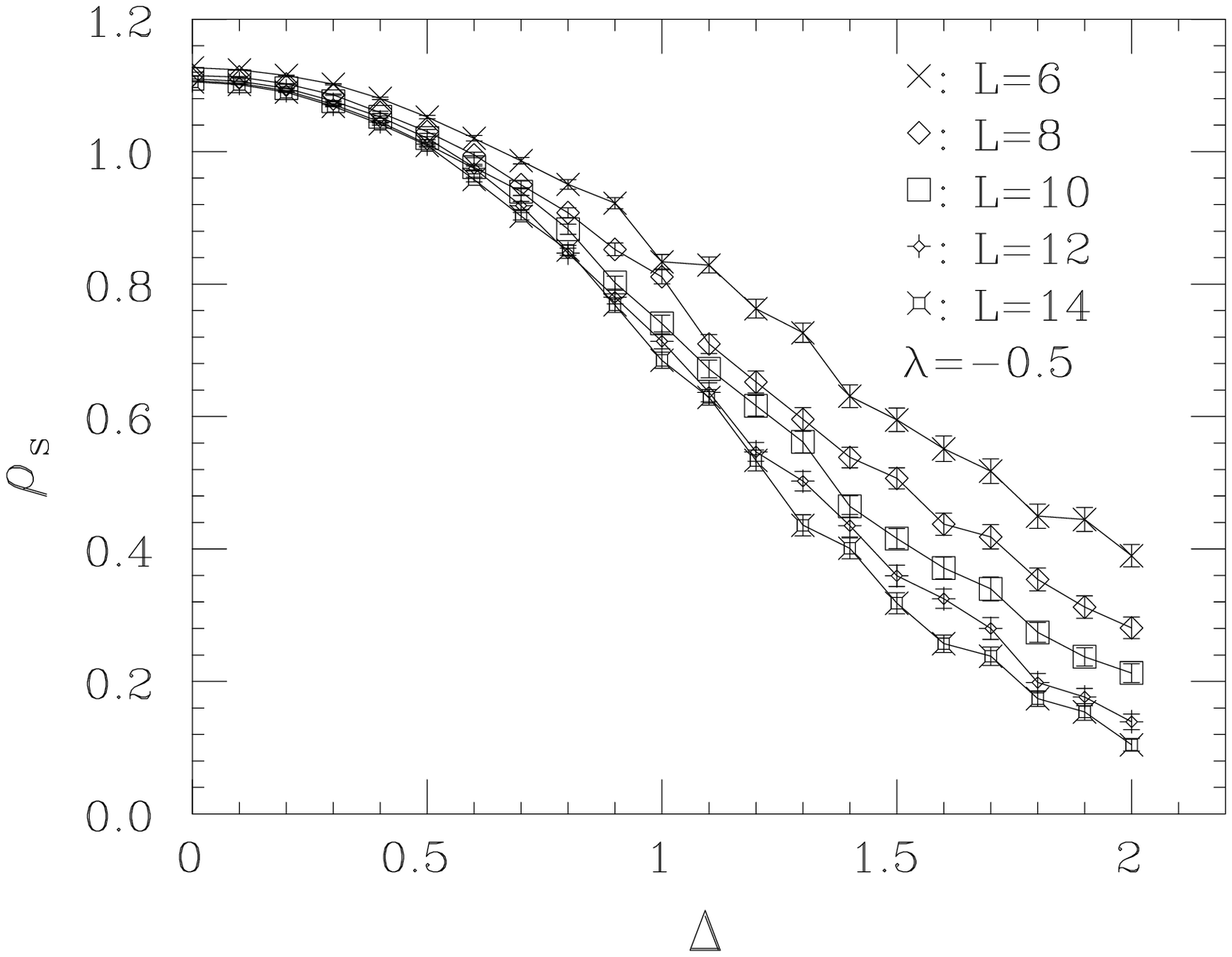}
\end{center}
\caption{
Plot of the spin stiffness ({\protect \ref{stiffness}}) 
against the randomness at $\lambda=-0.5$.}
\label{sf_mt5}
\end{figure}

\begin{figure}[htbp]
\begin{center}\leavevmode
\epsfxsize=17cm
\epsfbox{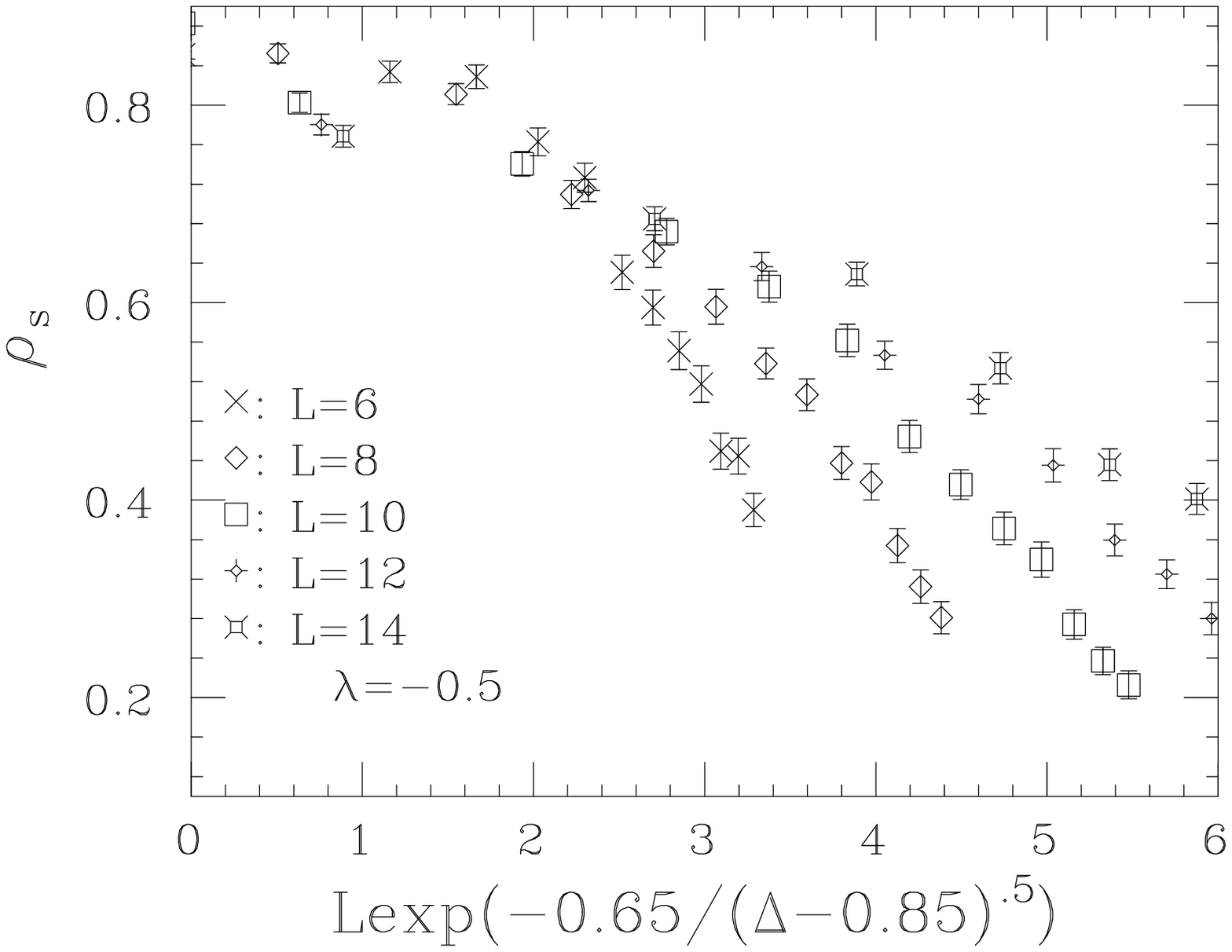}
\end{center}
\caption{
Scaling plot of the data shown in Fig. {\protect \ref{sf_mt5}}.
This plot yields the estimate $\Delta_{\rm c}=0.85\pm0.30$.}
\label{kt_scaling_mt5}
\end{figure}

\begin{figure}[htbp]
\begin{center}\leavevmode
\epsfxsize=17cm
\epsfbox{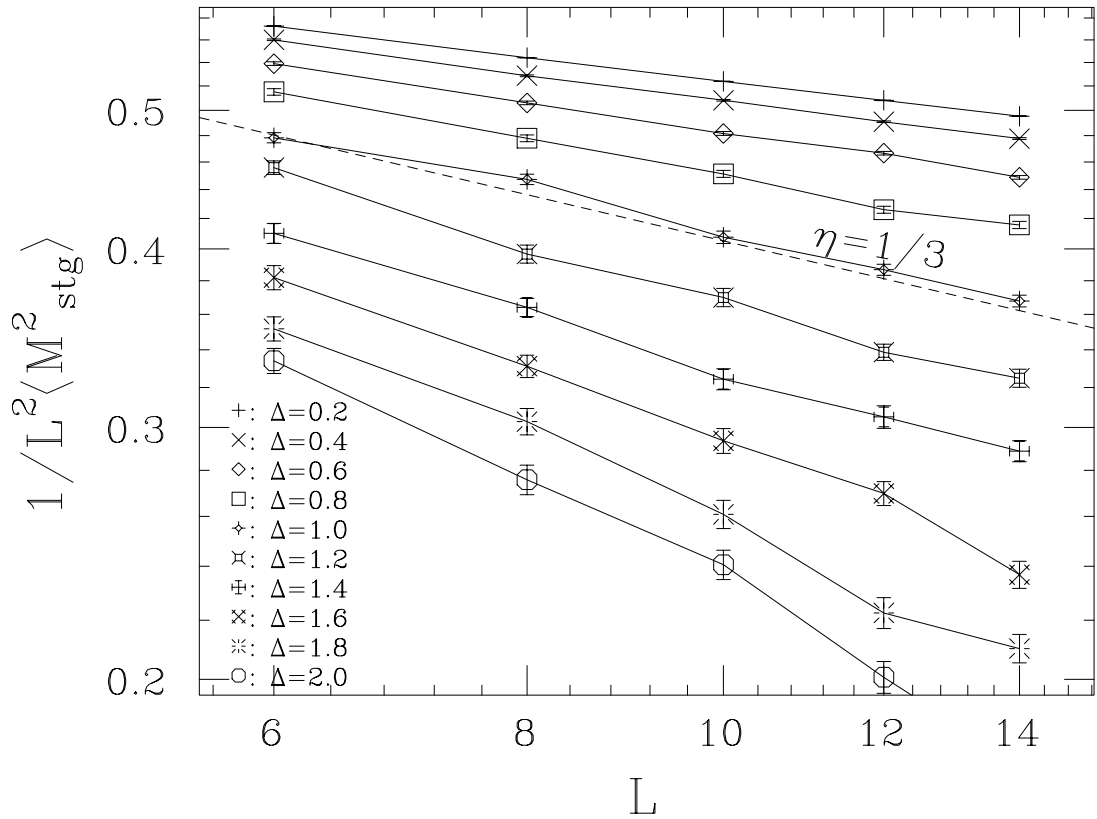}
\end{center}
\caption{
The square of the staggered magnetic moment ({\protect \ref{staggered}})
is plotted
for various randomnesses at $\lambda=-0.5$.
The dashed slope shows the decay rate $\eta_{\rm c}=1/3$,
which is proposed for the criterion for the 
localization-delocalization transition.}
\label{corr_mt5}
\end{figure}

\begin{figure}[htbp]
\begin{center}\leavevmode
\epsfxsize=17cm
\epsfbox{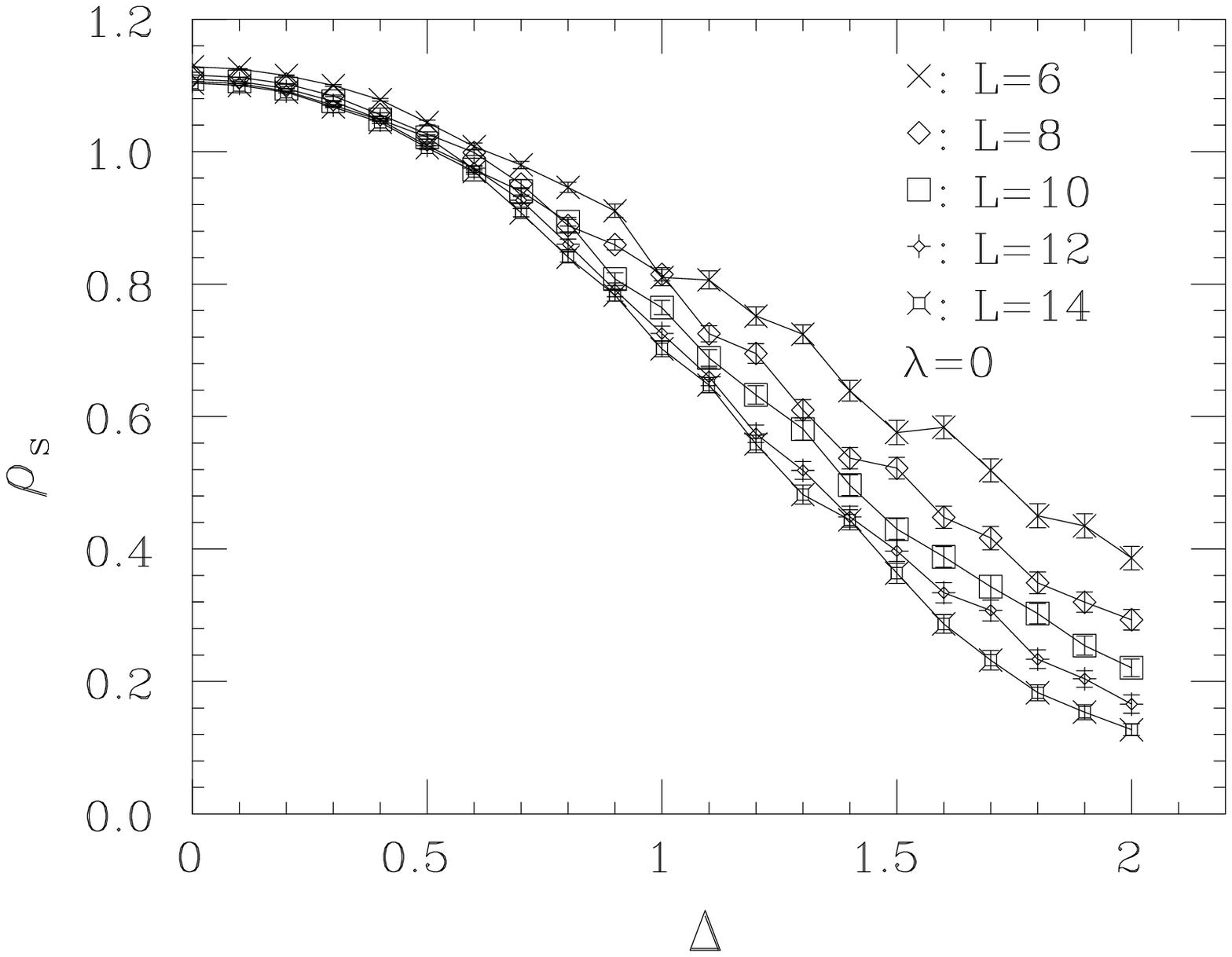}
\end{center}
\caption{
Plot of the spin stiffness ({\protect \ref{stiffness}}) 
against the randomness at $\lambda=0$.}
\label{sf_0}
\end{figure}

\begin{figure}[htbp]
\begin{center}\leavevmode
\epsfxsize=17cm
\epsfbox{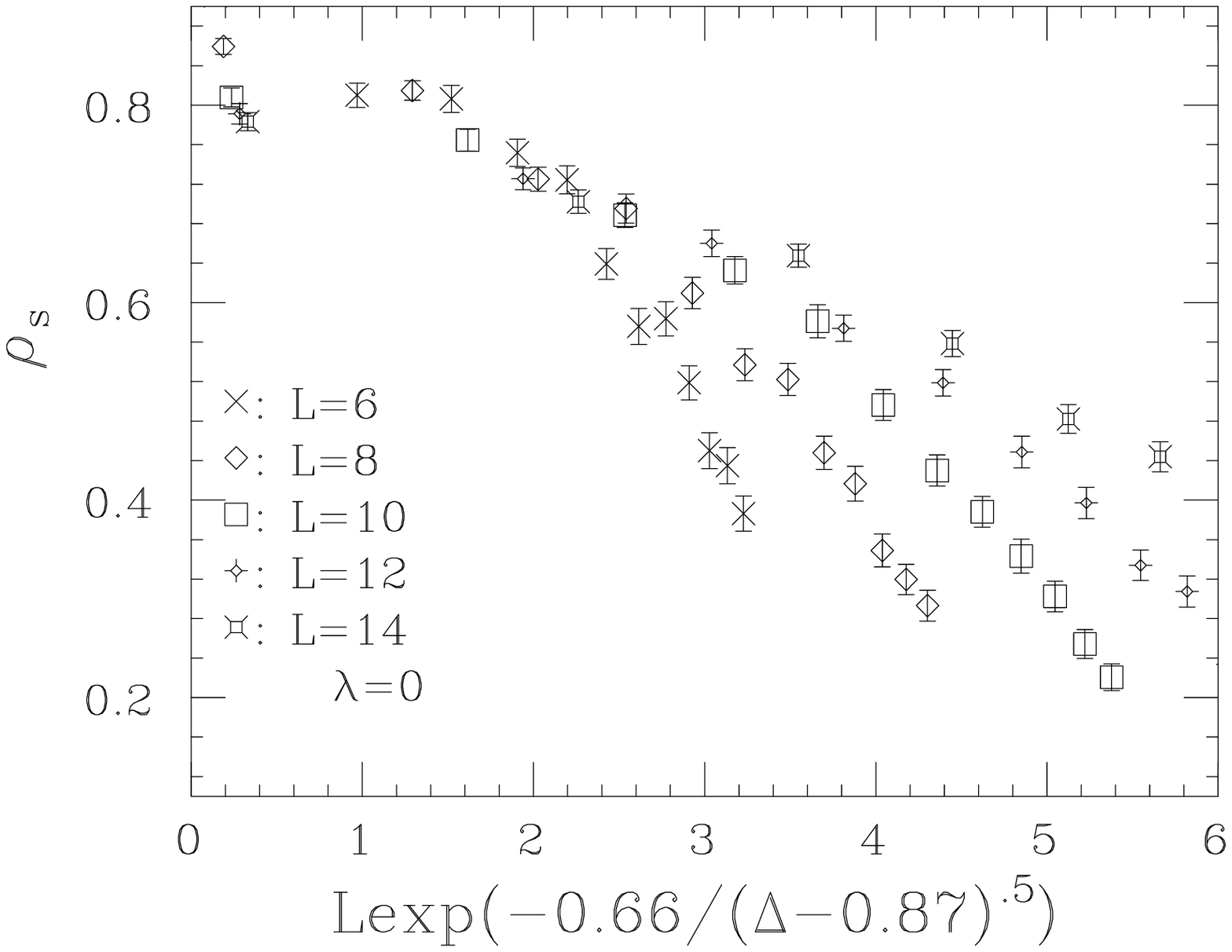}
\end{center}
\caption{
Scaling plot of the data shown in Fig. {\protect \ref{sf_0}}.
This plot yields the estimate $\Delta_{\rm c}=0.87\pm0.30$.}
\label{kt_scaling_0}
\end{figure}

\begin{figure}[htbp]
\begin{center}\leavevmode
\epsfxsize=17cm
\epsfbox{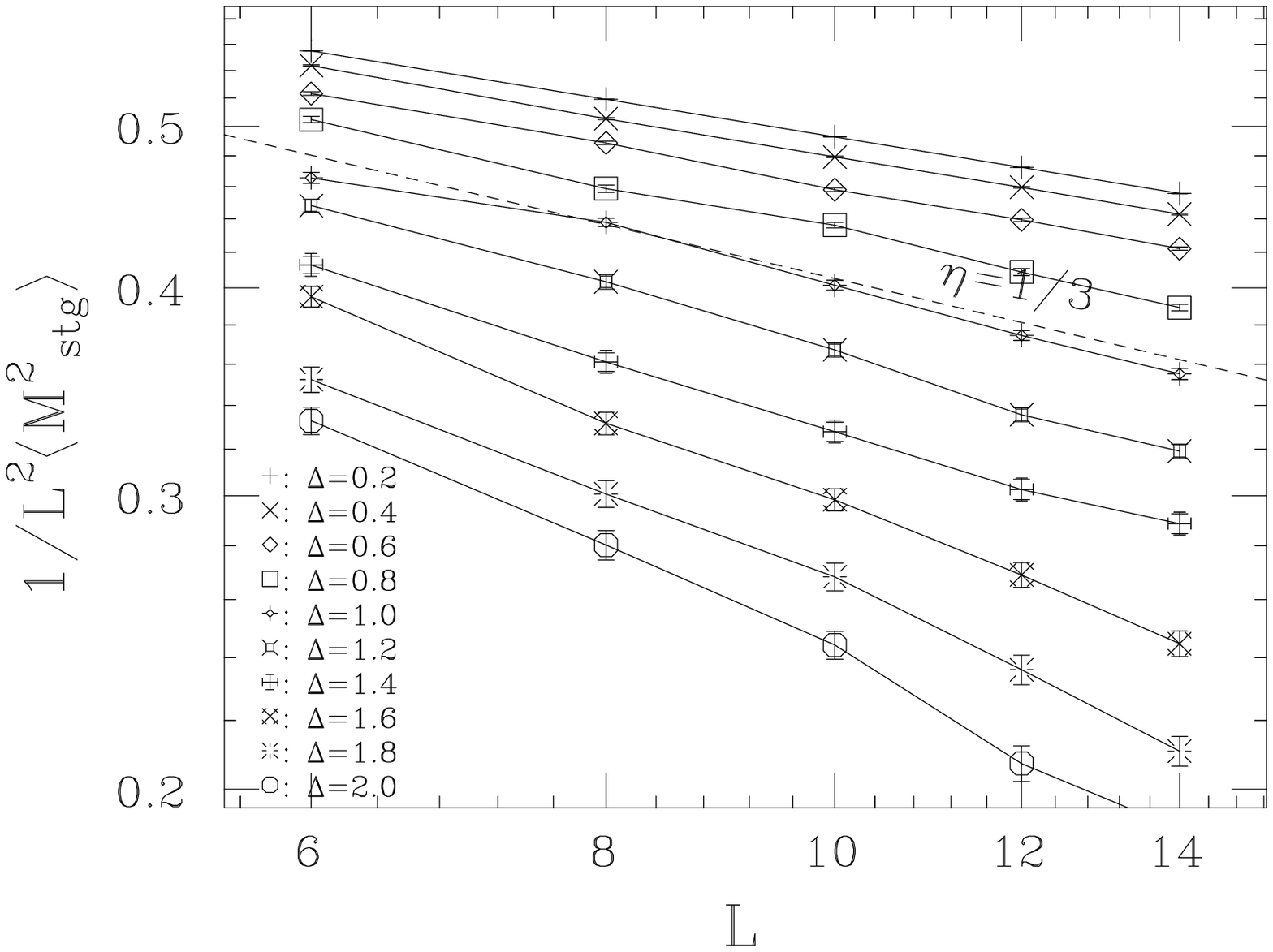}
\end{center}
\caption{
The square of the staggered magnetic moment ({\protect \ref{staggered}})
is plotted
for various randomnesses at $\lambda=0$.
The dashed slope show the decay rate $\eta_{\rm c}=1/3$.}
\label{corr_0}
\end{figure}

\begin{figure}[htbp]
\begin{center}\leavevmode
\epsfxsize=17cm
\epsfbox{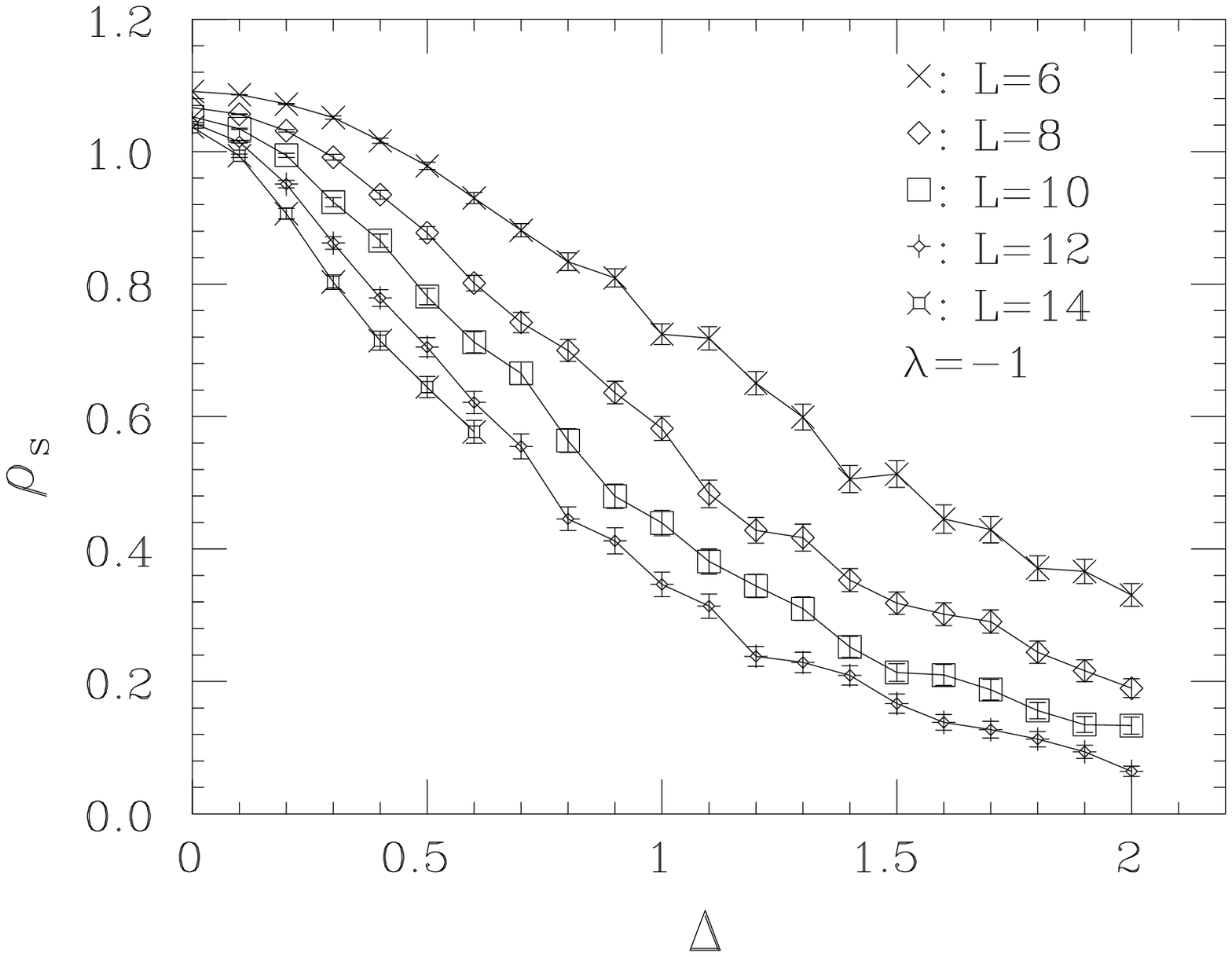}
\end{center}
\caption{
Plot of the spin stiffness ({\protect \ref{stiffness}}) 
against the randomness at $\lambda=-1$.}
\label{sf_m1}
\end{figure}

\begin{figure}[htbp]
\begin{center}\leavevmode
\epsfxsize=17cm
\epsfbox{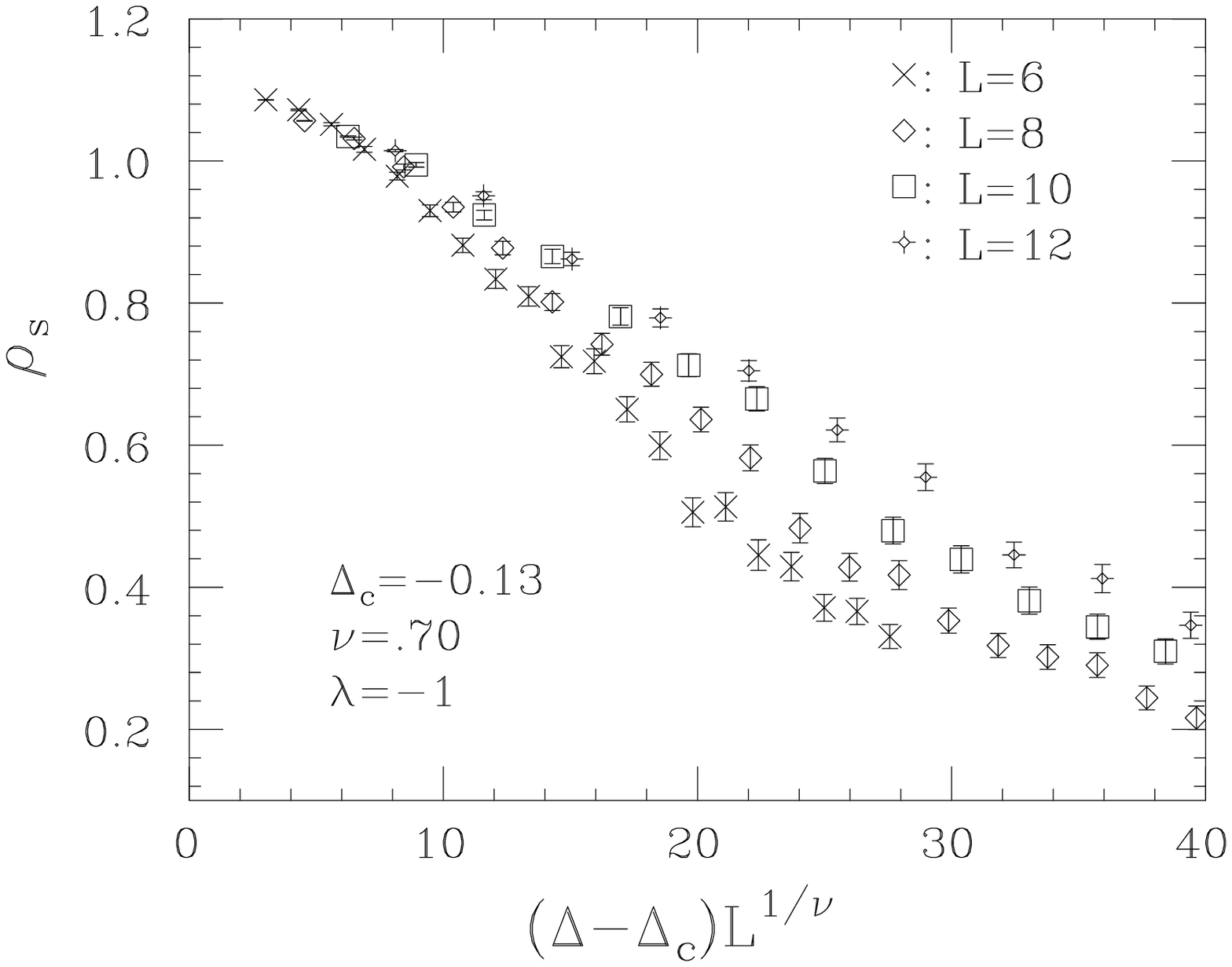}
\end{center}
\caption{
Scaling plot of the data shown in Fig. {\protect \ref{sf_m1}}.
This plot yields the estimate $\Delta_{\rm c}=-0.13\pm0.10$.}
\label{pw_scaling_m1}
\end{figure}

\begin{figure}[htbp]
\begin{center}\leavevmode
\epsfxsize=17cm
\epsfbox{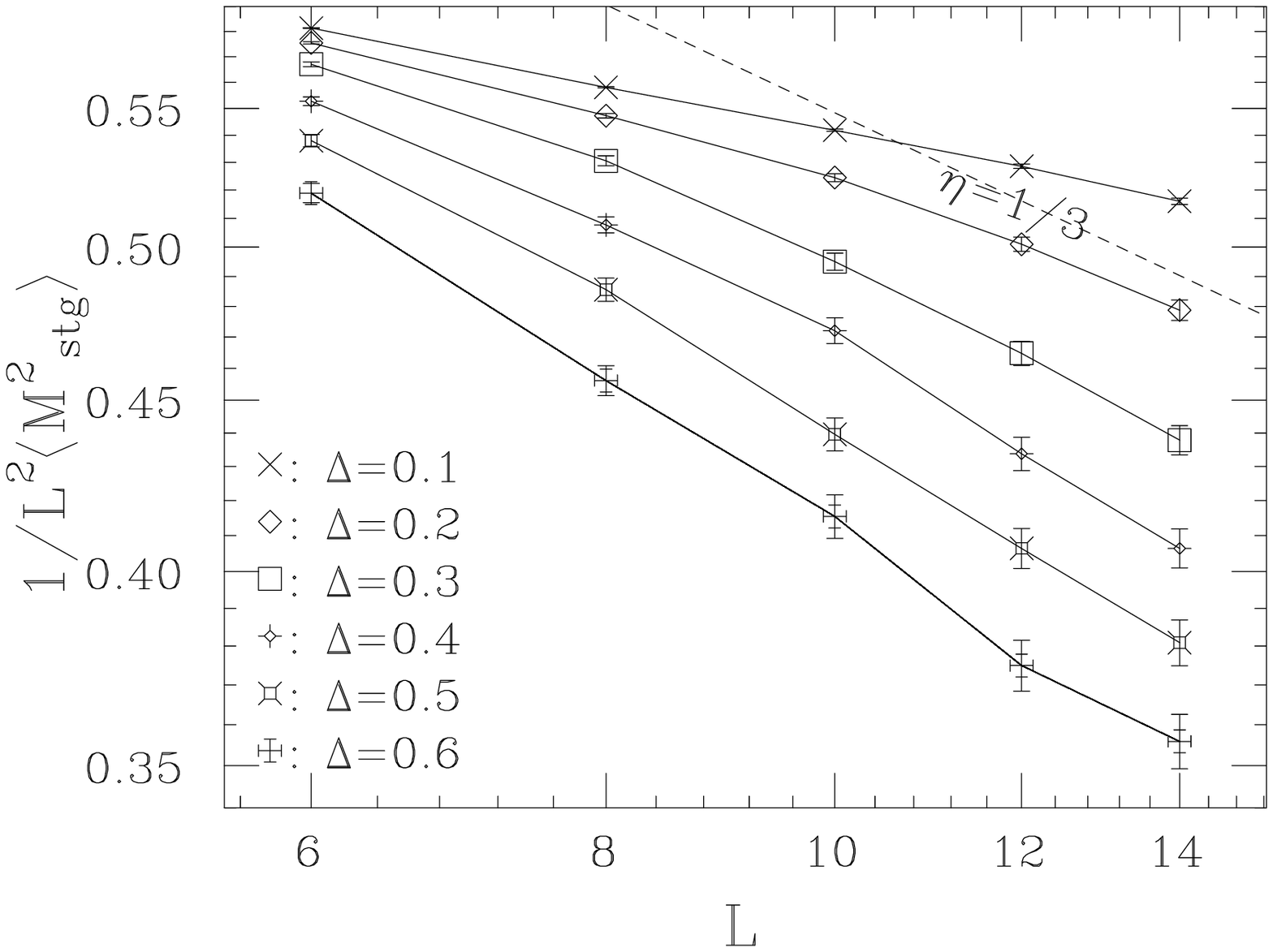}
\end{center}
\caption{
The square of the staggered magnetic moment ({\protect \ref{staggered}})
is plotted
for various randomnesses at $\lambda=-1$.
The dashed slope show the decay rate $\eta_{\rm c}=1/3$.}
\label{corr_m1}
\end{figure}

\begin{figure}[htbp]
\begin{center}\leavevmode
\epsfxsize=17cm
\epsfbox{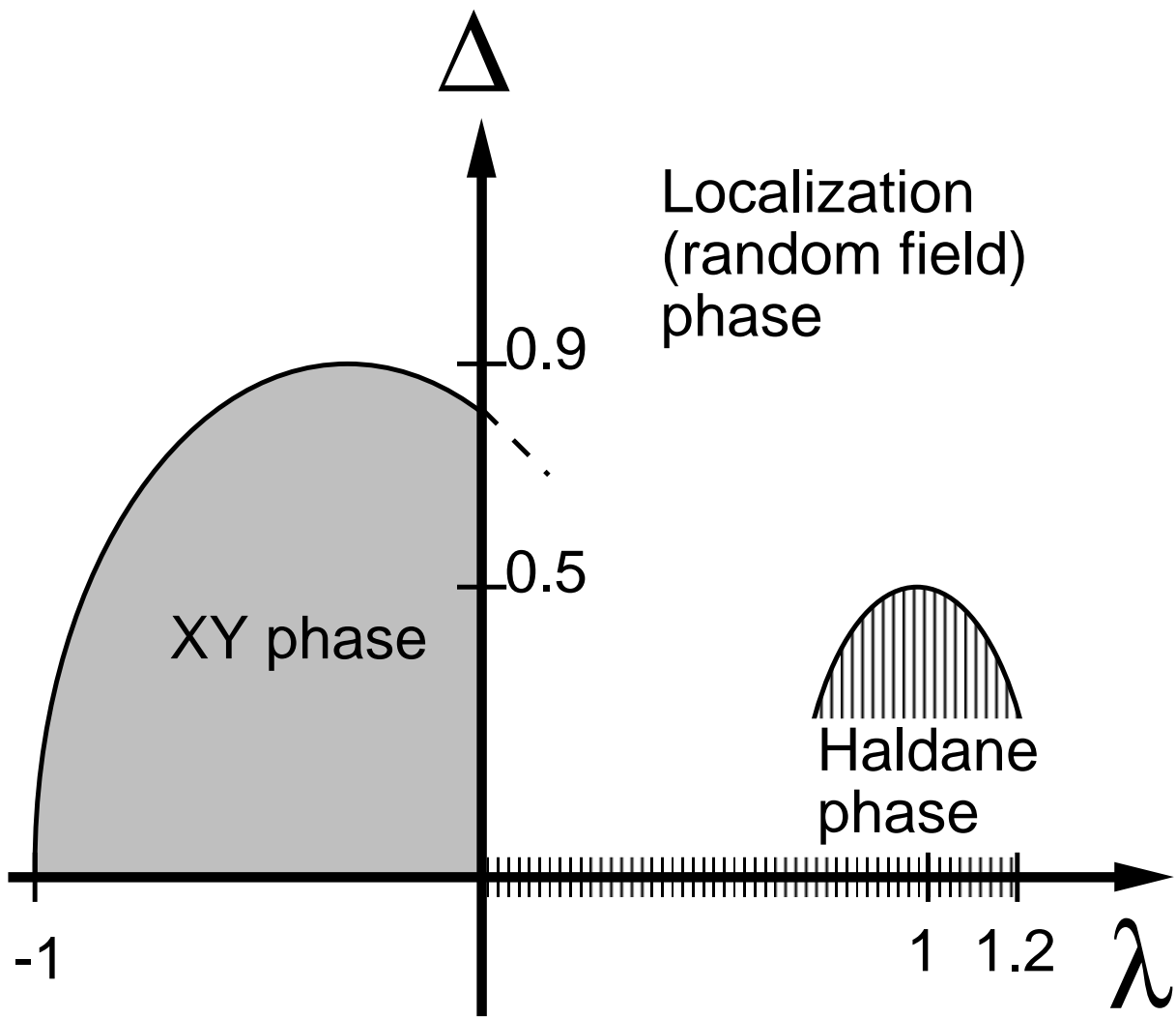}
\end{center}
\caption{
Phase diagram of the $S=1$ $XXZ$ model with the random
magnetic field;
the Hamiltonian is given by eq. ({\protect \ref{Hamiltonian}}).}
\label{phase_diagram_S1}
\end{figure}

\begin{figure}[htbp]
\begin{center}\leavevmode
\epsfxsize=17cm
\epsfbox{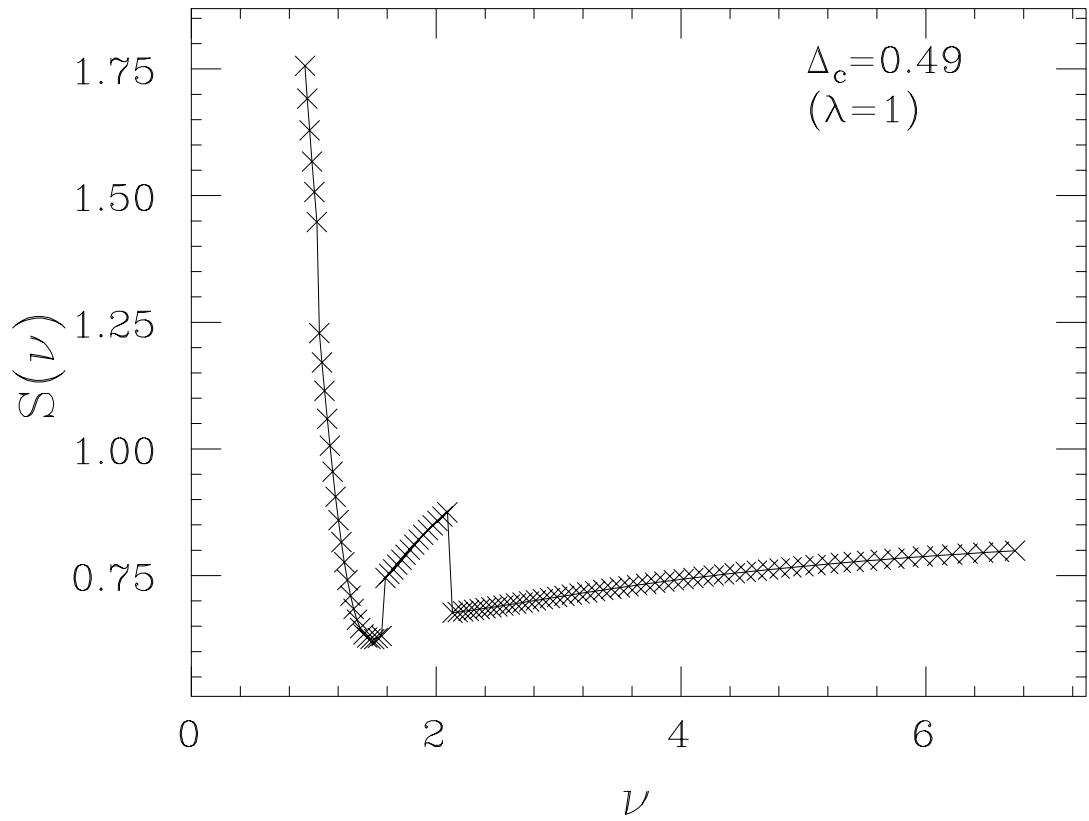}
\end{center}
\caption{
Local linearity function $S$ ({\protect \ref{linear}}) for the scaling data 
shown in Fig.
{\protect \ref{pw_scaling_1}} is plotted for various trial 
values of $\nu$ with
$\Delta_{\rm c}=0.49$ fixed.
The location of the minimum yields the estimate of the exponent $\nu%
\sim1.5$.}
\label{linearity}
\end{figure}

\end{large}
\end{document}